\documentclass[12pt]{iopart}
\usepackage{times}
\usepackage{graphicx}

\def\H{{\cal H}}
\def\L{{\cal L}}

\def\op#1{\hat{#1}}
\def\ket#1{| #1 \rangle}
\def\lket#1{| #1 \rangle\rangle}
\def\bra#1{\langle #1 |}

\def\vec#1{{\bf #1}}

\def\rmi{{\rm i}}

\begin{document}

\title{Dissipative `Groups' and the Bloch Ball}
\author{Allan I. Solomon\footnote{a.i.solomon@open.ac.uk} and Sonia G. Schirmer\footnote{sgs29@camb.ac.uk}}
\address{Quantum Processes Group, The Open University, Milton Keynes MK7 6AA}
\begin{abstract}
We show that a quantum control procedure on a two-level system
including dissipation  gives rise to  the semi-group
corresponding to the Lie algebra $gl(3,R)\oplus R^3$. The physical
evolution may be modelled by the action of this semi-group on a
3-vector as it moves inside the Bloch sphere, in the Bloch ball.
\end{abstract}

\section{Introduction}
Recent developments in quantum computing have emphasized the need
for a realistic analysis of dissipation in systems which have the
potential for use as qubits. In this note we discuss the effects
of control and dissipation on a two-level system. For a single
qubit {\em pure } state, it is well known that the unitary
evolution may be visualised as the movement of a vector, the Bloch
vector, on the surface of a 2-sphere, the Bloch sphere. In this
note we extend the idea to a two-level {\em mixed} state. For this
system, unitary evolution is on a spherical shell within the
Bloch Sphere. Dissipation causes more general motion within the
Bloch ball.   This motion corresponds to the action of a certain
{\em semi-group}.

 We also show that the effects of
dissipation may not be compensated by the interaction with the
external control. However, taking into account the effects of
 measurement, which may be modelled by certain projection
operators, the dissipative effects may indeed be modified,
allowing more effective control of the system.

\section{Dynamics of dissipative quantum control systems}
In pure-state quantum mechanics the state of the system is
usually represented by a wavefunction $\ket{\Psi}$, which is an
element of a Hilbert space $\H$.  For dissipative quantum
systems, however, a quantum statistical mechanics formulation is
necessary since dissipative effects can and do convert pure
states into statistical ensembles and vice versa.  In this case,
the state of the system is represented by a density operator
$\op{\rho}$, whose diagonal elements  determine the populations
of the energy eigenstates, while the off-diagonal elements
 determine the coherences between
energy eigenstates, which distinguish coherent superposition
states $\ket{\Psi}=\sum_{n=1}^N c_n \ket{n}$ from statistical
ensembles of energy eigenstates (i.e., mixed states) $\op{\rho}=
\sum_{n=1}^N w_n \ket{n}\bra{n}$. For a non-dissipative system
the time evolution of the density matrix $\op{\rho}(t)$ with
$\op{\rho}(t_0)=\op{\rho}_0$ is governed by
\begin{equation} \label{eq:rho_evol}
  \op{\rho}(t) = \op{U}(t) \op{\rho}_0 \op{U}(t)^\dagger,
\end{equation}
where $\op{U}(t)$ is the time-evolution operator satisfying the
Schrodinger equation
\begin{equation} \label{eq:SE}
  \rmi\hbar \frac{d}{dt}\op{U}(t) = \op{H}(\vec{f})\op{U}(t), \qquad \op{U}(0)=\op{I},
\end{equation}
where $\op{I}$ is the identity operator.  $\op{\rho}(t)$ also
satisfies the quantum Liouville equation
\begin{equation} \label{eq:LE}
  \rmi\hbar \frac{d}{dt}\op{\rho}(t) = [\op{H}(\vec{f}),\op{\rho}(t)]
  = \op{H}(\vec{f})\op{\rho}(t) - \op{\rho}(t)\op{H}(\vec{f}).
\end{equation}
$\op{H}(\vec{f})$ is the total Hamiltonian of the system, which
depends on a set of control fields $f_m$:
\begin{equation} \label{eq:H_expansion}
  \op{H}(\vec{f}) = \op{H}_0 + \sum_{m=1}^M f_m(t) \op{H}_m,
\end{equation}
where $\op{H}_0$ is the internal Hamiltonian and $\op{H}_m$ is
the interaction Hamiltonian for the field $f_m$ for $1\le m\le M$.

The advantage of the Liouville equation (\ref{eq:LE}) over the
unitary evolution equation (\ref{eq:rho_evol}) is that it can
easily be adapted to dissipative systems by adding a dissipation
(super-)operator $\L_D[\op{\rho}(t)]$:
\begin{equation} \label{eq:dLE}
   \rmi\hbar\dot{\rho}(t) = [\op{H}_0,\op{\rho}(t)] +
   \sum_{m=1}^M f_m(t) [\op{H}_m,\op{\rho}(t)] + \rmi\hbar\L_D[\op{\rho}(t)].
\end{equation}

In general, uncontrollable interactions of the system with its
environment lead to two types of dissipation: phase decoherence
(dephasing) and population relaxation (decay). The former occurs
when the interaction with the enviroment destroys the phase
correlations between states, which leads to a decay of the
off-diagonal elements of the density matrix:
\begin{equation} \label{eq:dephasing}
 \dot{\rho}_{kn}(t)
  = -\frac{\rmi}{\hbar}([\op{H}(\vec{f}),\op{\rho}(t)])_{kn}-\Gamma_{kn}\rho_{kn}(t)
\end{equation}
where $\Gamma_{kn}$ (for $k\neq n$) is the dephasing rate between
$\ket{k}$ and $\ket{n}$. The latter happens, for instance, when a
quantum particle in state $\ket{n}$ spontaneously emits a photon
and decays to another quantum state $\ket{k}$, which changes the
populations according to
\begin{equation} \label{eq:poptrans}
\dot{\rho}_{nn}(t) =
-\frac{\rmi}{\hbar}([\op{H}(\vec{f}),\op{\rho}(t)])_{nn}
  +\sum_{k\neq n} \left[\gamma_{nk}\rho_{kk}(t)-\gamma_{kn}\rho_{nn}(t)\right]
\end{equation}
where $\gamma_{kn}\rho_{nn}$ is the population loss for level
$\ket{n}$ due to transitions $\ket{n}\rightarrow\ket{k}$, and
$\gamma_{nk}\rho_{kk}$ is the population gain caused by
transitions $\ket{k}\rightarrow\ket{n}$.  The population
relaxation rate $\gamma_{kn}$ is determined by the lifetime of
the state $\ket{n}$, and for multiple decay pathways, the
relative probability for the transition
$\ket{n}\rightarrow\ket{k}$.   Phase decoherence and population relaxation lead to a
dissipation superoperator (represented by an $N^2 \times N^2$
matrix) whose non-zero elements are
\begin{equation}
  \begin{array}{ll}
  (\L_D)_{kn,kn} = -\Gamma_{kn} & k \neq n \\
  (\L_D)_{nn,kk} = +\gamma_{nk} & k \neq n \\
  (\L_D)_{nn,nn} = - \sum_{n\neq k} \gamma_{kn}.
 \end{array}
\end{equation}

Population decay and dephasing allow us to overcome kinematical
constraints such as unitary evolution to create statistical
ensembles from pure states, and pure states from statistical
ensembles, which is important for many applications such as
optical pumping.  However, there are instances when this is not
desirable such as in quantum computing, where these effects
destroy quantum information.  Thus, there are situations when we
would like to prevent decay and dephasing.  A cursory glance at
the quantum Liouville equation for coherently driven, dissipative
systems (\ref{eq:dLE}) suggests that it might be possible to
prevent population and phase relaxation by applying suitable
control fields such that
\begin{equation}
   \sum_{m=1}^M f_m(t) [\op{H}_m,\op{\rho}(t)]+\rmi\hbar\L_D[\op{\rho}(t)] = 0.
\end{equation}
Unfortunately, however, a more careful analysis reveals that this
is \emph{not} possible, in general, as we shall now show
explicitly for a two-level system, or qubit in quantum computing
parlance.

\section{Dynamics of a 2-level system subject to control, decay and dephasing}

The Hamiltonian for a driven two-level system with energy levels
$E_1<E_2$ is
\begin{equation}
  \op{H}[\vec{f}(t)]  = \op{H}_0 + f_1(t) \op{H}_1 + f_2(t) \op{H}_2
\label{fam}
\end{equation} where $\op{H}_0$ is the internal
Hamiltonian and $\op{H}_1$ and $\op{H}_2$ represent interaction
Hamiltonians with independent (real-valued) control fields
$f_1(t)$ and $f_2(t)$,
\[
  \op{H}_0 =    \left[ \begin{array}{cc} E_1 & 0 \\ 0 & E_2 \end{array} \right], \;
  \op{H}_1 = d_1\left[ \begin{array}{cc} 0 & 1   \\ 1 & 0   \end{array} \right], \;
  \op{H}_2 = d_2\left[ \begin{array}{cc} 0 & -\rmi\\ \rmi & 0 \end{array} \right].
\]
$d_1$, $d_2$ are the (real-valued) dipole moments for the
transition and $\omega= (E_2-E_1)/\hbar$ is the transition
frequency.

We can re-write the Liouville equation in matrix form in a higher
dimensional space, often referred to as Liouville space.
Straightforward computation shows that
\begin{equation}
   \frac{d}{dt}\lket{\rho(t)}=\L
   \lket{\rho(t)}
  = [(1/{\rmi\hbar)}(\L_0 + f_1(t) \L_1 + f_2(t) \L_2) + \L_D] \lket{\rho(t)}
\end{equation}
where $ \lket{\rho(t)} =(\rho_{11}(t), \rho_{12}(t),\rho_{21}(t) , \rho_{22}(t))^{T}$ \\
\[
 \L_0 =
  \left( \begin{array}{cccc}
          0 & 0 & 0 & 0 \\
          0 & -\hbar\omega & 0 & 0 \\
          0 & 0 & +\hbar\omega & 0 \\
          0 & 0 & 0 & 0
          \end{array} \right)
 \; \; \; \; \L_1 = d_1
   \left( \begin{array}{cccc}
           0  & -1 & +1 & 0 \\
           -1 & 0  &  0 & +1 \\
           +1 & 0  &  0 & -1 \\
           0  & +1 & -1 & 0
          \end{array} \right)
\]
\[
  \L_2 = d_2\left( \begin{array}{cccc}
          0  & -\rmi & -\rmi & 0 \\
          +\rmi & 0  &  0 & -\rmi \\
          +\rmi & 0  &  0 & -\rmi \\
          0  & +\rmi & +\rmi & 0
         \end{array} \right)
   \; \; \; \; \L_D =\left( \begin{array}{cccc}
          -\gamma_{21} & 0 & 0 & \gamma_{12} \\
          0 & -\Gamma & 0 & 0 \\
          0 & 0 & -\Gamma & 0 \\
          \gamma_{21} & 0 & 0 & -\gamma_{12}
          \end{array} \right)
\label{liou}
\] $\gamma_{12}$ is the rate of population relaxation
from $\ket{2}$ to $\ket{1}$, $\gamma_{21}$ is the rate of
population relaxation from $\ket{1}$ to $\ket{2}$ (usually zero),
and $\Gamma$ is the dephasing rate.

Notice that the matrix elements of the Liouville operators $\L_1$
and $\L_2$ are zero where the matrix elements of the dissipation
operator $\L_D$ are non-zero, and vice versa.  Thus, no matter
how we choose the control fields, we cannot cancel the effect of
the dissipative terms.  The best we can do is to use coherent
control to implement quantum error correction schemes to restore
decayed/dephased quantum states to their
correct values.  An early contribution to the theory of continuous feedback for such  systems is contained in \cite{wiseman}, while  a more recent scheme of continuous quantum error correction  involving weak measurements and feedback has  been proposed in \cite{ahn}.

\section{Dynamical Semi-group and the Bloch Ball}
The more usual real vector form for $\rho$ is $\rho_B \equiv
(\rho_{1,2}+\rho_{2,1},i(\rho_{1,2}-\rho_{2,1}),\rho_{1,1}-\rho_{2,2},\rho_{1,1}+\rho_{2,2}).$
In the general case this is referred to as the {\em coherence
vector}\cite{lendi}. Since for the systems under consideration we
shall take $\rho_{1,1}+\rho_{2,2}$ as constant (no population
loss) only the first three components of $\rho_B$ transform under
the dynamics. For {\em pure states} the norm is constant - thus
generating motion on the surface of a 2-sphere, the Bloch
sphere.   For our more general scenario, motion takes place in
the interior of this sphere, the {\em Bloch ball}.

The family of control hamiltonians generate the Lie
algebra $u(2)$, in this case a completely controllable
system\cite{SFS2001}. The Lie algebra generated by the matrices
corresponding to Eq.(\ref{liou}) acting on $\rho_B$ is the
inhomogeneous algebra $gl(3,R)\oplus R^3$. The evolution of the
system in time is determined by $\exp(\L t)$. Noting that the
eigenvalues of $\L_D$ are (always) negative, the demand that our
set of operators remain bounded gives in effect a {\em
semi-group}, with only unlimited {\em positive} values of $t$
permitted.  These considerations may be generalized to any
dimensions \cite{solsch}.

\section{Conclusions}

In this note we showed how the effects of control and dissipation
can be treated geometrically, by the movement of the coherence
vector inside the Bloch ball.  We incidentally noted that
dissipative effects could not be compensated by control dynamics alone.
Without measurements and feedback, dissipation usually forces the system  into an equilibrium state.  One can  see that without any control this state corresponds to a point on the z-axis of the Bloch sphere; with constant controls one can show that the state  converges to a point on an ellipse inside the Bloch ball.  The coordinates of these attractors can easily be expressed as a function of the dissipative terms.

Intuitively,  error correction is not possible with 
  control fields  alone   because the dissipative terms tend to pull us inside the Bloch ball and, 
  without quantum measurement  feedback, we can't get back to the surface of the ball because 
  the control fields  can only perform rotations.

The analysis  can be treated by traditional Lie algebraic
methods, by use of a dynamical Lie algebra.  In the specific case
of a qubit treated here, this algebra is the inhomogeneous
semi-direct sum $gl(3,R)\oplus R^3$. The control fields generate a rotation algebra within this larger algebra. Imposing boundedness
conditions on the evolution of the dynamics obtained by
exponentiation of this algebra leads to a semi-group description
of the evolution.

Although due to restrictions of space we have treated explicitly only the case of a two-level system, the methods can be generalised without difficulty to any finite level system.  For an $N$-level system, the coherence vector is essentially an $N^2-1$ real component vector (for population-preserving dynamics) whose motion is restricted to the $N^2-2$-dimensional Bloch sphere for non-dissipative systems, and to the interior of the corresponding Bloch ball in general. The  associated inhomogeneous real algebra is given by semi-direct sum  $gl(N^2-1,R)\oplus R^{(N^2-1)}$, and this gives rise   
 by means of exponentiation to the corresponding semi-group which determines the dynamics.  Finally, for {\em quasi-spin} systems, defined by symmetric population decay rates, the inhomogeneous (translation-like) terms disappear.

\end{document}